\documentclass[final,5p,times,twocolumn]{elsarticle}
\usepackage{amssymb}
\usepackage{amsmath}
\usepackage{booktabs}
\usepackage{amssymb}
\usepackage{graphicx}
\usepackage{color} 
\usepackage{subfig}

\journal{PLOS One}

\begin{document}
\begin{frontmatter}
\title{Bcl-2 family controls mitochondrial outer membrane permeabilization by performing non-trivial pattern recognition.}
\author{Tomas Tokar, Jozef Ulicny}
\address{Department of Biophysics, University of P. J. Safarik in Kosice, Jesenna 5, 040 01, Kosice, Slovakia}
\begin{abstract}
Interactions between individual members of the B-cell lymphoma 2 (Bcl-2) family 
of proteins form a regulatory network governing mitochondrial outer membrane 
permeabilization (MOMP). Bcl-2 family initiated MOMP causes release of 
the inter-membrane pro-apoptotic proteins to cytosol and creates a cytosolic 
environment suitable for the executionary phase of apoptosis.
Using mathematical modeling and computational simulation, we have analyzed 
the response of this regulatory network caused by the up-/downregulation of 
each individual member of the Bcl-2 family. 
As a result, a non-linear stimulus-response emerged, the characteristics of 
which are associated with bistability and switch-like behavior.
Using the principal component analysis (PCA) we have shown that the Bcl-2 
family classifies the random combinations of inputs into two distinct classes, 
and responds to these by one of the two qualitatively distinct outputs.
As we showed, the emergence of this behavior requires specific organization of 
the interactions between particular Bcl-2 proteins.
\end{abstract}

\begin{keyword}
apoptosis, Bcl-2, bistability, pattern recognition, mitochondria
\end{keyword}

\end{frontmatter}

\section*{Introduction}

Programmed cell death (PCD), often denoted as a cellular suicide, plays 
an important role in the homeostasis of every multi-cellular organism
~\cite{ouyang_2012}. 
One of the main forms of PCD is called apoptosis~\cite{wyllie_2010,ulukava_2011,
ouyang_2012}, a process which is well distinguished by its characteristic 
morphology~\cite{elmore_2007}.
Defects in apoptosis regulation may cause a variety of serious diseases, 
including the neurodegenerative disorders~\cite{mattson_2006}, 
autoimmune diseases~\cite{nagata_autoimmune_2010}, or cancer
~\cite{burz_apoptosis_2009,fulda_2009,strasser_deciphering_2011}.
Apoptosis can be initiated by either extracellular stimuli or by signals
originating from a cell's internal space~\cite{strasser_apoptosis_2000,
strasser_deciphering_2011}.
Signals initiating apoptosis then proceed through the apoptotic signaling 
pathways, which contain several control points
~\cite{strasser_apoptosis_2000,strasser_deciphering_2011}.
One of the most important of such points, integrating a multitude of incoming
apoptotic (and antiapoptotic) signals is formed by a family of Bcl-2 
(B-cell lymphoma 2) proteins~\cite{danial_2004, chipuk_2008}.
The Bcl-2 family controls mitochondrial outer membrane 
permeabilization (MOMP)~\cite{tait_2010,landes_2011}, the crucial 
event of apoptosis.

MOMP allows the release of key apoptotic players - Smac/DIABLO and 
cytochrome c, from a mitochondrial intermembrane space to cytosol
~\cite{tait_2010,landes_2011}.
In the presence of ATP, released cytochrome c binds to a cytosolic protein Apaf-1, 
causing Apaf-1 oligomerization and recruitment of an inactive pro-caspase-9, 
leading to formation of a multi-protein complex known as an apoptosome 
~\cite{mace_2010,perez-paya_2011,kulikov_2012}.
Within the apoptosome, pro-caspase-9 subsequently undergoes processing and 
activation~\cite{mace_2010,perez-paya_2011,kulikov_2012}.
The active caspase-9 proteolytically activates caspase-3~\cite{wurstle_2012}.
Smac/DIABLO, once released to the cytosol, inhibits XIAP 
(X-linked inhibitor of apoptosis) - the most prominent suppressor
of caspases-3 and -9~\cite{martinez-ruiz_2008}.
Caspase-3 and other effector caspases (caspases-6 and -7) 
are the primary executioners of the apoptosis
~\cite{olsson_2011, strasser_deciphering_2011}. 
Activation of effector caspases signifies the point of no-return, 
after which apoptosis irreversibly occurs~\cite{green_1998}.

Bcl-2 family members are functionally classified as either antiapoptotic, 
or proapoptotic. Structurally, Bcl-2 proteins can be categorized according to 
the number of Bcl-2 homology domains (BH) in their $\alpha$-helical regions
~\cite{elkholi_2011,strasser_deciphering_2011}.
Antiapoptotic members (Mcl-1, A1, Bcl-xL, Bcl-2, Bcl-w and Bcl-B) 
are characterized by the presence of four BH domains (BH1-4)
~\cite{chipuk_reunion_2010,placzek_2010}.
Their role is to prevent MOMP by inhibition of proapoptotic family members
~\cite{chipuk_reunion_2010,placzek_2010}.
Proapoptotic members can be
divided to BH3-only proteins and multidomain proteins - effectors
~\cite{strasser_deciphering_2011}. 
BH3-only proteins can be further subdivided based on their 
role in apoptotic signaling. 
BH3-only subgroup members, termed sensitizers, or enablers
(Noxa, Bad, Puma, Hrk, Bmf and Bik) can only bind to 
antiapoptotic Bcl-2 proteins, forming inactive dimers
~\cite{elkholi_2011}.
Members of another BH3 subgroup, termed activators (Bim and Bid), 
can act in the same way~\cite{elkholi_2011}, 
but in addition, activators can directly activate effectors
~\cite{chipuk_reunion_2010,westphal_2011}.
Effectors, once activated, undergo oligomerization and form pores --
mitochondrial apoptosis channels (MAC) in 
the mitochondrial outer membrane (MOM), leading eventually to MOMP.
~\cite{dejean_2010,landes_2011}.
Therefore, effectors are the primary target of inhibition by their 
antiapoptotic relatives~\cite{westphal_2011}.

Individual Bcl-2 family members are regulated by wide-variety of factors, 
e.g. growth factor deprivation, cytokine withdrawal, heat shock, 
DNA damage, hypoxia, death receptors stimulation and many others
~\cite{chipuk_reunion_2010}. 
Mechanisms of regulation include transcription control and/or
post-translational modifications by phosporylation, or proteolytic cleavage
~\cite{chipuk_reunion_2010}.
Bcl-2 family thus integrates a multitude of converging signals to decide 
whether to commit MOMP or not.
This decision is carried in an all or nothing manner, 
giving no possibility of intermediate MOMP.
This interesting behavior has made the Bcl-2 family an 
attractive subject of mathematical modeling and computer simulations.

There are several works regarding modeling and simulation of 
the Bcl-2 family and the control of MOMP, 
revealing and examining a variety of non-linear system behaviors such as 
robustness, stimulus-response ultrasensitivity
~\cite{chen_robustness_2007} 
and steady-state bistability~\cite{cui_two_2008,sun_2010,tokar_2012}. 
Besides these, the Bcl-2 family was involved in several other, 
more general models of apoptosis signaling
~\cite{bagci_2006,albeck_quantitative_2008,harrington_2008}.

In the above mentioned works, the authors reduce the complexity of their model 
by grouping of several functionally similar species together.
Usually, the Bcl-2 family's members are assigned into four groups 
according to their structural and functional classification. 
The most prominent group's member is taken as the model's representation of 
the whole group of species. 
Previous models of the Bcl-2 regulatory network differ by level of details, 
nevertheless they all adopt such simplification.
Although, such simplification provides an attractive trade-off between 
the model's complexity and plausibility, functional specificities 
of Bcl-2 family's individuals are being omitted.

In the proposed work we provide a literature-based mathematical model 
in which interactions between individual Bcl-2 family members are 
distinguished.
Our goal was to investigate the behavior of the detailed model, in particular 
to prove/disprove the switching properties obtained by models based on 
functional grouping. 
In the process we obtained additional insight into the decision mechanism of 
the Bcl-2 control of MOMP.
The non-trivial pattern-recognition emerged as a consequence of functional 
specificities of Bcl-2 family individuals.
In addition, an explicit model of pair interactions allowed us to probe the pro- and anti-apoptotic potency of individual
members of the Bcl-2 family and to rank them according to their ability to promote or prevent the MOMP event.

\section*{Results}

\subsection*{Bistability of the Bcl-2 family regulation of MOMP}
Several previous works
~\cite{cui_two_2008,sun_2010,tokar_2012} have been focused on the analysis 
of bistable behavior of the Bcl-2 family mediated regulation of MOMP.
Similarly, we performed a variation of the single stimulus 
parameter of the model to analyze steady state stimulus-response 
of the model.
The production rate of the tBid ($\mbox{kptBid}$) was considered as 
the stimulus and the steady state concentration of the MAC 
($\mbox{MAC}_{\mbox{ss}}$) was the measure of response.
The stimulus -- $\mbox{kptBid}$ was varied through the chosen range 
of values, while the other parameters of the model remained unchanged. 

\begin{figure*}[!ht]
\begin{center}
\includegraphics[width=0.5\textwidth]
{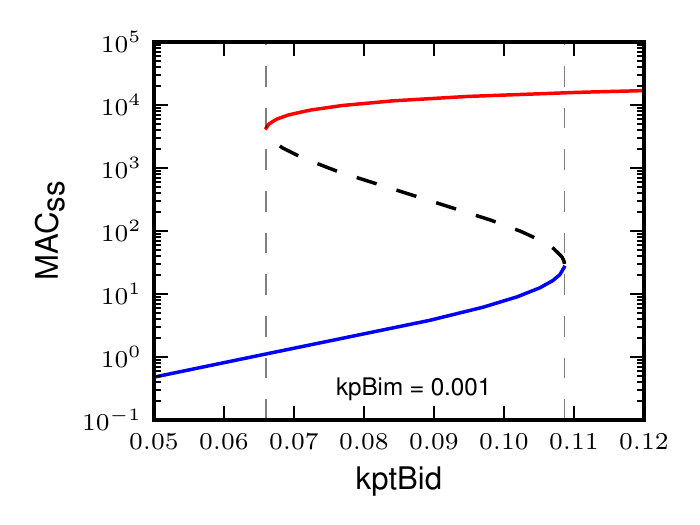}
\end{center}
\caption{
{\bf Steady state concentration of the MAC ($\mbox{MAC}_{\mbox{ss}}$) is plotted as 
a function of the production rate of tBid (kptBid).}
$\mbox{MAC}_{\mbox{ss}}$ is increasing with increasing value of 
$\mbox{kptBid}$. 
The $\mbox{MAC}_{\mbox{ss}}$ remains at very low levels (pro-survival -- the blue solid curve), 
until the production rate exceeds the threshold 
(right vertical dashed line).
Exceeding the threshold value causes sudden increase of the 
$\mbox{MAC}_{\mbox{ss}}$ (onset of MOMP -- red solid curve). 
The subsequent decrease of the production of tBid cause only 
slow decrease of $\mbox{MAC}_{\mbox{ss}}$, until the another threshold is 
crossed (left vertical dashed line).
Then the $\mbox{MAC}_{\mbox{ss}}$ suddenly drops back to very low levels.
Vertical dashed lines are enclosing the bistable region. 
Within this region system can persist in one of the two stable steady states
(solid curves), which are separated by unstable steady states (dashed curve).}
\label{fig:bistability_tbid}
\end{figure*}

Using the parameters of the model listed in the 
Table~\ref{tab:params} we have obtained the stimulus-response
dependence depicted in the figure
~\ref{fig:bistability_tbid}.
The obtained steady state stimulus-response forms hysteresis, the typical hallmark 
of bistability of the dynamic system. 
The two thresholds (Fig~\ref{fig:bistability_tbid}, marked by the 
left and right vertical dashed lines) enclose the bistable region. 
Within the bistable region, the system under the same value of the 
stimulus may occur in one of two stable steady states, depending 
on the initial conditions. These steady states are separated by the 
unstable steady states (marked by dashed curve). 
Such systems are often described as  
the mechanism of a ``toggle switch"\cite{tyson_2003}.


Previous works limited themselves to the response of the Bcl-2 family 
induced by activators tBid/Bim solely.
In the presented work we have explored the model's response to
variation of all components, including both activators, all the 
anti-apoptotic proteins, enablers and effectors.
Utilizing individual production rates as the input stimuli, we performed 
set of analyses analogous to the previous one. 

\begin{figure*}[!H]
\centering
\begin{center}
\subfloat[]{\includegraphics[width=0.4\textwidth]
{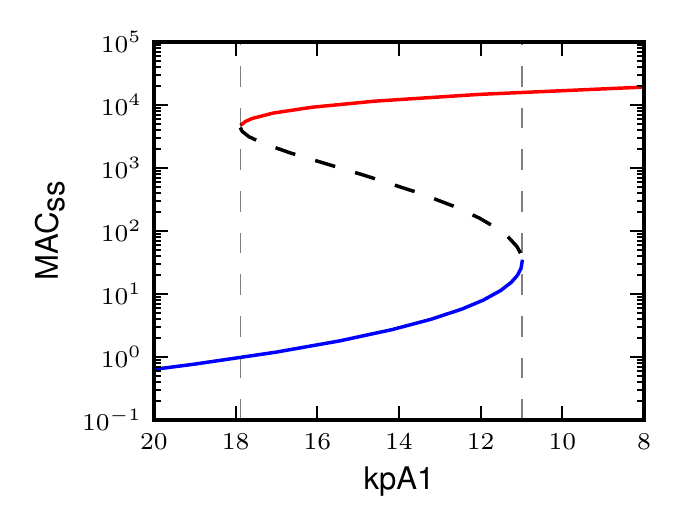}}
\subfloat[]{\includegraphics[width=0.4\textwidth]
{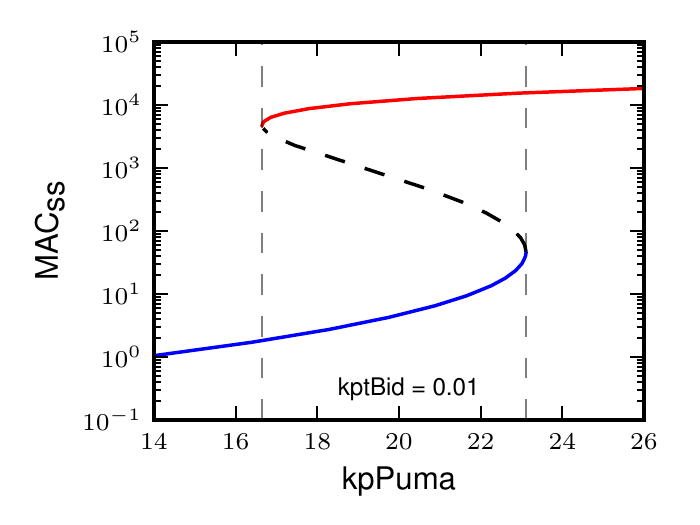}} \\
\subfloat[]{\includegraphics[width=0.4\textwidth]
{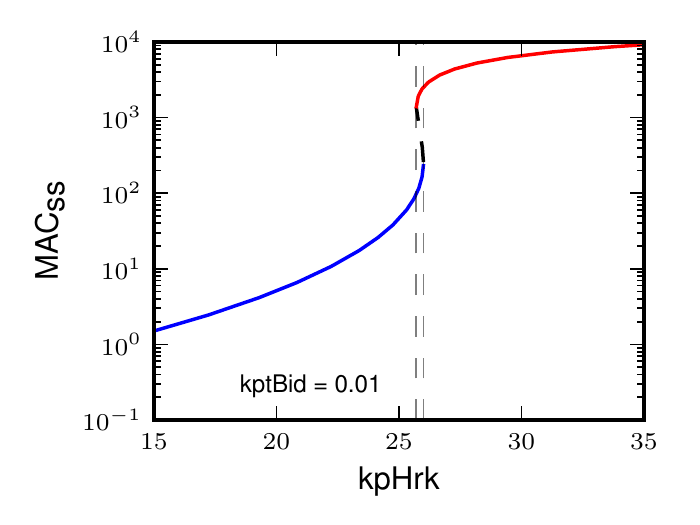}}
\subfloat[]{\includegraphics[width=0.4\textwidth]
{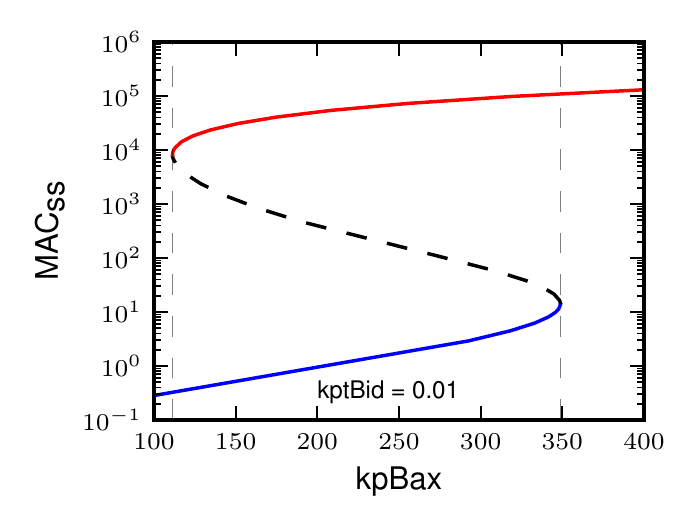}}
\end{center}
\caption{
{\bf Steady state concentration of the MAC plotted as 
a function of indicated production rates.}}
\label{fig:bistability_others}
\end{figure*}

In addition to tBid, we have observed that steady state stimulus-response 
hysteresis resulted also from stimulation by the second of activators -- Bim 
(data not shown), as well as by the the anti-apoptotic proteins.  
Surprisingly, even a variation of the production rates of enablers Puma and Bad \& Bmf yields 
stimulus-response hysteresis (see Fig~\ref{fig:bistability_others}).

Variation of the Hrk \& Bik productions yields hysteresis as well, but the obtained
range of bistability is extremely narrow, close to ultrasensitive sigmoid curve. 
Similarly to bistability, sigmoid curve means that modeled system is 
insensitive to low levels of the given input, but it can respond significantly
to its high levels. 
In contrast to bistability, sigmoidal response is not discretized, as the
response is continuously increasing/decreasing with the growing/reducing 
input strength. While the bistability can be compared to mechanism of 
``toggle switch" sigmoid stimulus-response is often compared to the
functioning of the ``push-button"~\cite{tyson_2003}.

Robust hysteresis with a wide bistable region was yielded by the variation of 
production rate of Bax and less pronounced hysteresis was obtained by variation 
of production of Bak.
This shows that ``toggle'' switching of the Bcl-2 response can also be achieved
by significant upregulation of the effector proteins.

\begin{figure*}[!H]
\begin{center}
\includegraphics[width=0.5\textwidth]
{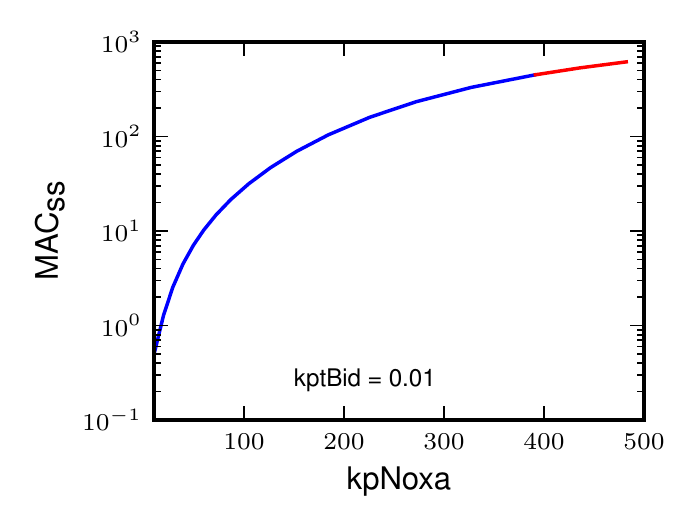}
\end{center}
\caption{
{\bf Steady state concentration of the MAC is plotted as 
a function of the production rate of Noxa.}
Hyperbolic curve indicates response sensitivity to event a small
levels of stimulation.}
\label{fig:hyperbolic_noxa}
\end{figure*}

Interestingly, variation of the production rate of the Noxa produced
a hyperbolic response. Hyperbolic dependence, in contrast to hysteresis 
and sigmoid response, indicates sensitivity to even a small increase of 
input stimuli. Therefore, stimulation of the model through the production of 
Noxa can ``tune'' the model's sensitivity 
to other incoming stimuli (see Fig~\ref{fig:hyperbolic_noxa}).

\subsection*{Monte-Carlo variation of production rates results in bimodal 
	     distribution of steady-state abundance of MAC}
In the previous section, we analyzed the dependence of the steady state 
concentration of MAC on the production of individual proteins. 
In what follows, we performed Monte-Carlo analysis of the dependence of the 
steady state concentration of MAC on the simultaneous variation of multiple 
production rates.

In each iteration, the values of all the production rates 
(kpMcl1, kpA1, kpBclXl, kpBcl2, kpBclw, kpBclB, kpHrk, kpBik, kptBid, kpPuma, 
kpBim, kpBad, kpBmf, kpNoxa, kpBax, kpBak) of the model were 
varied according to following rule:
\begin{equation}\label{eq:variation_rule}
  kp = kp^{\ast} \cdot 10^{q}\,
\end{equation}
where $kp$ is the variated production rate, $kp^{\ast}$ is its default value
and $q$ is the random, uniformly distributed real number chosen from 
$\langle-2, 2\rangle$ interval. 
Other parameters of the model were kept at their default values.
Similar to the analysis of bistability, the steady state concentration of MAC,
was used as the output.
 
\begin{figure*}[!H]
\begin{center}
\includegraphics[width=0.5\textwidth]
{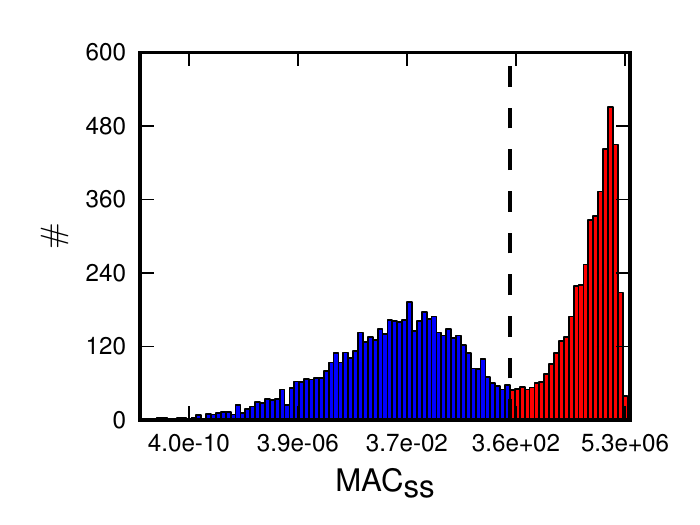}
\end{center}
\caption{
{\bf Distribution of steady state concentrations of MAC produced by
$10^4$ random variations of the model's production rates.}
Vertical dashed line denotes the minimum between two local maxima, 
defining the threshold value of the steady state concentration of the MAC, 
distinguishing pro-survival and pro-MOMP responses.}
\label{fig:MACss_hist}
\end{figure*}

We have plotted the distribution of the model's response obtained for $10^4$ 
iterations.
The results (see Fig.~\ref{fig:MACss_hist}) show clear bimodal distribution of 
the model's response.
The bimodal distribution of the response proves that the Bcl-2 regulatory 
network can turn random combinations of inputs into two qualitatively 
distinct outputs
~\cite{gardner_2000,bagowski_2001,ozbudak_2004,sun_2010}.

The minimum located between the local maxima of the response distribution
(Fig~\ref{fig:MACss_hist}, marked by vertical dashed line) was considered as 
the threshold value of steady state concentration of MAC, 
separating the pro-survival (colored blue) and MOMP (colored red) responses.
Hereafter, the steady state concentrations of the MAC below this threshold
are considered as pro--survival, while the concentrations above the 
threshold are considered as MOMP initiating (pro--MOMP).
It is worth mentioning, that the value of the threshold intersects 
all the hysteresis curves 
(Fig~\ref{fig:bistability_tbid} and~\ref{fig:bistability_others}), 
within the bistable range.
The value of this threshold -- $\sim 350$, assumes that MOMP occurs once 
the number of effector dimers in mitochondria surrounding environment exceeds 
350.
This is remarkably good agreement with experimental estimations made by 
Martinez-Caballero et al.~\cite{martinez_caballero_2009}.

\subsection*{Bcl--2 family performs non-trivial pattern recognition.}
Using the same sets of the production rates as were used in 
the previous analysis we arranged the matrix of stimuli. 
Each row of this matrix -- stimuli vector corresponds to one iteration of 
the Monte-Carlo analysis from the previous section and each of its columns
corresponds to the one of the production rates, defining the size of the 
stimuli matrix to $10^{4}\times16$. Each column was normalized to its mean.
Then we performed the principal component analysis (PCA) of the matrix of stimuli 
and plotted the input vectors within the plane defined by principal components.

\begin{figure*}[!H]
\begin{center}
\subfloat[]{
\includegraphics[width=0.4\textwidth]{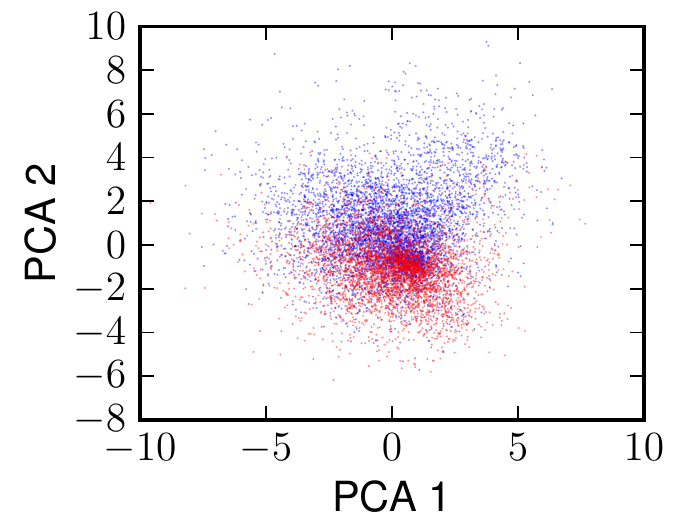}}
\subfloat[]{
\includegraphics[width=0.4\textwidth]{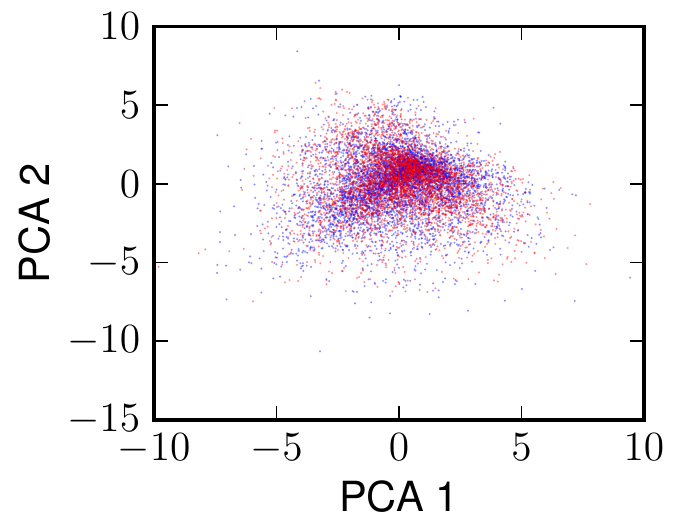}}
\end{center}
\caption{
{\bf Scatter plot of all input combinations, plotted in the 
plane defined by the principal components analysis.}
Inputs associated with pro-MOMP response are colored red, remaining 
inputs are colored blue. The clusterization of 
inputs according to response quality is apparent for reference model (a),  
but obviously absent when altering the model's topology (b).}
\label{fig:pca}
\end{figure*}

The results (see Fig~\ref{fig:pca}, left) show that when the random input stimuli 
vectors are plotted in the PCA--defined plane, they form a randomly scattered 
cloud. 
But, when the each vector is colored based on the associated response,
it appears that the stimuli associated with the given response are
clustered.
This shows, that the Bcl-2 regulatory network is capable of 
taking a wide range of random combinations of incoming signals and 
classifying them into two sharply defined responses of distinct biological relevance.
Such functionality defines what is in the field of machine learning and neural
networks known as non-trivial pattern recognition
~\cite{bray_1995,haykin_1999,jain_2000,helikar_2008}.

In following, we created ten alternative models of the Bcl-2 
family, by mutating the topology of the default model. 
By mutation we mean random addition of non-existing or deletion of 
existing inhibitory interactions between anti-apoptotic and pro-apoptotic
proteins and/or activations of effectors by activators. 
Such mutations allow alteration of the Bcl-2 family model on its detailed level, 
but preserve the interactions between the functional groups consistent
with the default model.

For each of the alternative models we have performed a total of five of such 
mutations. We then performed Monte-Carlo analysis of these models by 
generating the $10^4$ random combinations of input stimuli. 
For each alternative model we then identified the threshold value of the 
effectors activity and subsequently performed the PCA of the stimuli matrix.

As a result we have found that all the alternative models produced bimodal 
distribution of response, but none of them clustered input stimuli similarly to 
the default model (see Fig~\ref{fig:pca}, right side).
This indicate that, while the bistability can emerge from alternative
topologies of the Bcl-2 family interaction network, 
the pattern recognition is strictly associated with this particular topology
of this regulatory network. 

\subsection*{Pivotal pro- and anti-apoptotic Bcl-2 family members}
In the following we wanted to answer the naturally arising question: 
Which of the production rates are pivotal regarding the determination of 
the model's response? 

The model's response varies over several orders of magnitude. 
However, all the response values below/above the threshold, 
regardless of the value itself, are considered to be qualitatively equal -- 
pro--survival/pro--MOMP, providing the same biological consequences.
Therefore, we were interested in correlation of the values of the production 
rates with the response quality (pro--survival/pro--MOMP), instead of the 
correlation with the response quantity. 

We utilized the point-biserial correlation coefficient (PBCC) - $r_{pb}$ as the 
the measure of the correlation between the value of the production rate and
quality of the model's response (pro--survival/ pro--MOMP).
PBCC is frequently used to measure the correlation between the two variables,
one of which is dichotomous (either naturally, or artificially dichotomized)
~\cite{tate_1954}.
PBCC for each production rate was calculated as follows:
\begin{equation}
r_{pb} = \frac{M  - S}
{\sigma}\sqrt{\frac{n_M\cdot n_S}{n^2}}\,, \\
\end{equation}
where the $M$ and $S$ are the means of values of the given production rate
corresponding to pro--MOMP and pro--survival responses respectively.
$n_M$ and $n_S$ are the number of values of the given production rate, 
corresponding to pro--MOMP and pro-survival responses respectively.
$n$ is the total number of the values of the given production rate and 
$\sigma$ is its standard deviation.

\begin{figure*}[!H]
\begin{center}
\includegraphics[width=0.6\textwidth]{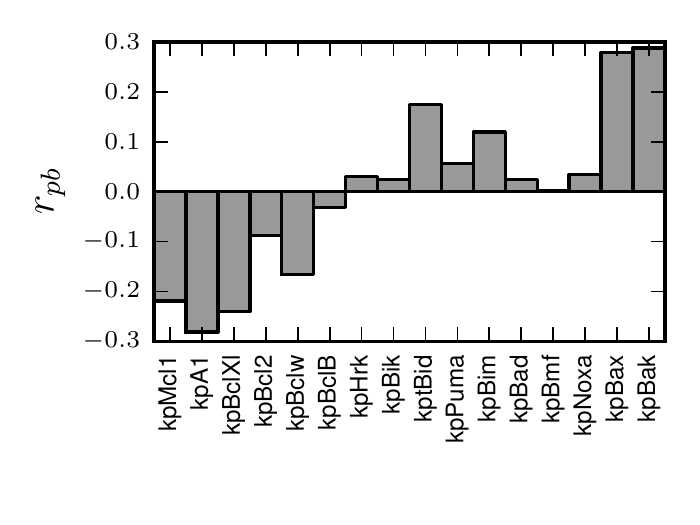}
\end{center}
\caption{{\bf Point-biserial correlation coefficients ($r_{pb}$) as the measure of correlation between the values of the production rates
and the model's response.}}
\label{fig:pbccs}
\end{figure*}

The results we obtained (figure~\ref{fig:pbccs}) prove
the role of the tBid and Bim -- the only activators of effectors -- as the
primary pro-apoptotic proteins. 
Similarly, the Puma is the most efficient MOMP promoter among 
the enabler proteins.
On the other hand, the most efficient MOMP preventers are proteins 
A1 and Bcl-Xl, followed by Mcl-1 and Bcl-w.

\section*{Discussion}
The Bcl-2 family of proteins consists of sixteen
(excluding tissue-specific Bcl-2 ovarian killer -- Bok) proteins that differ in their upstream
regulation as well as in their interactions with other family members.
We translated current biological knowledge of these interactions into
mathematical model, which was utilized to study the regulation of the
MOMP -- crucial apoptotic event, which is maintained by the Bcl-2 family.

Analysis of the model's response to variation of productions of individual
proteins, revealed that the system property called steady-state bistability
emerges as a robust feature of the modeled system.
Our results indicate that the most of the Bcl-2 proteins can serve as 
bistable ``toggles", up- or downregulation of these cause 
``switching on'' the MOMP.
Steady-state bistability is currently being
,,the favorite framework for thinking about the switch between life and death"
~\cite{spencer_2011}. Therefore, the above-mentioned results were expected.
However, we have also found that other Bcl-2
proteins act as a ``push-buttons'' -- alternative switching
mechanisms with very narrow or missing bistable range and
one of them as a ``tuner'' of the model's sensitivity -- upregulation of which 
results in hyperbolic model's response.

We have shown that the model is able to process random combinations of inputs
to produce output that has bimodal distribution.
This strongly suggests that orchestration of these ``toggles'', ``push-buttons'' and
``tuners'' constitutes a molecular device whose function is to integrate
multitude of incoming, continuous inputs into binary outcomes.
Bimodal distribution was previously observed by Sun et al~\cite{sun_2010}
in the flow cytometry of Bax activation as the response to staurosporine
treatment of the HeLa cells population, supporting our finding.

Moreover, our results further suggest that this molecular device
can perform pattern recognition -- which is non-trivial functionality,
often associated with neural networks and machine learning algorithms.
This functionality is impaired by deletion of a relatively small
number of reactions, as well as by addition of artificial interactions,
even for interactions which are consistent with the relationships between
the functional groups of the family.

Measurements of the correlation between the individual signals and the model's
response predict that the most potent inputs of this network
are associated with the regulation of activators tBid and Bim, enabler protein
Puma and the anti-apoptotic sentinels A1, Bcl-Xl, Bcl-w and Mcl-1.

Finally, to outline the directions in which the proposed work could be extended,
we would like to point out the necessity of utilization of more sophisticated,
machine-learning based approaches to better analyze the synergy of the Bcl-2
family regulation of MOMP. We believe that its understanding can be crucial in
development of novel anti-cancer drugs and/or treatment of the serious
neurodegenerative diseases.

\section*{Model and its biological relevance}
\label{materials_and_methods}
In the proposed work, we modeled the MOMP regulatory network 
formed by the interactions of the members of the Bcl-2 family of proteins
(see Table~\ref{tab:family}).
Our model of the Bcl-2 family is defined by the extensive set of 
reactions which are listed in Table~\ref{tab:reactions}
and the respective set of reaction rates~\ref{tab:params},
which were estimated in accordance with the 
previously published coarse-grained models of Bcl-2 family.

In each simulation, the initial abundance of each protein was either 
set to zero for a protein that is not produced from external sources 
(reactions 46--61), otherwise defined by following rule:
\begin{equation}
a(t_0) = \frac{kp}{kdeg},
\end{equation}
setting the initial abundance to balance the production
(reactions 46--61) and degradation (catch-all reaction 62) of the given protein at the
initial conditions.
Default values of the reaction rates (listed in the Table
~\ref{tab:reactions}) were estimated
in accord with previously published models of the Bcl-2 family
~\cite{chen_robustness_2007,cui_two_2008,sun_2010,tokar_2012}. 
The abundance of the mitochondrial apoptosis channels -- MACs of 
the detailed model was calculated as the sum of the abundances of 
the aBax$\sim$aBax, aBak$\sim$aBak, aBax$\sim$aBak dimers.

Protein interactions were modeled using mass action kinetics, 
translated into a set of ordinary differential equations. 
The model was implemented and all the analyses were performed within 
Python programming language, by using the PySCes 
(Python Simulator for Cellular Systems)~\cite{olivier_2005} module.

\section*{Acknowledgments}
This work was supported by the Slovak Research and Development Agency
(contracts: APVV-0242-11, SEPO I ITMS code 26220120024 and SEPO II ITMS code 
26220120039)
and Scientific Grant Agency of the Ministry of Education of Slovak Republic 
under the grant VEGA-1/4019/07. Authors appreciate this support.

\clearpage
\bibliography{bibliography}
\bibliographystyle{elsarticle-num}
\clearpage
\section*{Figure Legends}

\clearpage
\newpage
\section*{Tables}

\begin{table*}[!ht]
\footnotesize
\center
\caption{
\bf{Interactions between individual members of the Bcl-2 family of proteins.}}
\begin{tabular}{llll}
\hline
\textbf{Group \&} & \textbf{Binds to \& inhibits:} & \textbf{Activates:} &\textbf{Ref}.\\
\textbf{Member} & & & \\
\hline
& & & \\
\textit{Antiapoptotic} & & & \\
\textit{proteins:} & & & \\
Mcl-1 & Noxa, Bim, Puma, Bax, Bak & & ~\cite{chen_differential_2005,strasser_deciphering_2011} \\
Bcl-2 & Bad, Bim, Puma, Bmf, Bax & & ~\cite{chen_differential_2005,strasser_deciphering_2011} \\
A1 & Noxa, Bim, Puma, Bid, Hrk, Bik, Bax, Bak & & ~\cite{chen_differential_2005,strasser_deciphering_2011} \\
Bcl-xL & Bad, Bim, Puma, Bid, Hrk, Bmf, Bik, Bak, Bax & & ~\cite{chen_differential_2005,strasser_deciphering_2011} \\
Bcl-w & Bad, Bim, Puma, Bid, Hrk, Bmf, Bik, Bax & & ~\cite{chen_differential_2005,strasser_deciphering_2011} \\
Bcl-B & Bax & & ~\cite{rautureau_2010}\\
& & & \\
\textit{Enablers:} & & & \\
Noxa & Mcl-1, A1 & & ~\cite{chen_differential_2005,elkholi_2011,strasser_deciphering_2011} \\
Bad  & Bcl-xL, Bcl-w, Bcl-2 & & ~\cite{chen_differential_2005,elkholi_2011,strasser_deciphering_2011} \\
Puma & Bcl-xL, Bcl-w, Bcl-2, Mcl-1, A1  & & ~\cite{chen_differential_2005,elkholi_2011,strasser_deciphering_2011} \\
Hrk & Bcl-xL, Bcl-w, A1 & & ~\cite{chen_differential_2005} \\
Bmf & Bcl-xL, Bcl-w, Bcl-2 & & ~\cite{chen_differential_2005,elkholi_2011} \\
Bik & Bcl-xL, Bcl-w, A1 & & ~\cite{chen_differential_2005} \\
& & & \\
\textit{Activators:} & & & \\
Bim & Bcl-xL, Bcl-w, Bcl-2, Mcl-1, A1 & Bax, Bak & ~\cite{chen_differential_2005,elkholi_2011,strasser_deciphering_2011} \\
tBid & Bcl-xL, Bcl-w, A1 & Bax, Bak & ~\cite{chen_differential_2005,elkholi_2011} \\
& & & \\
\textit{Effectors:} & & & \\
Bak & Bcl-xL, Mcl-1, A1 & & ~\cite{elkholi_2011,strasser_deciphering_2011} \\
Bax & Bcl-xL, Bcl-w, Bcl-2, Bcl-B, Mcl-1, A1 & & ~\cite{elkholi_2011,strasser_deciphering_2011} \\
\hline
\end{tabular}
\begin{flushleft}
\end{flushleft}
\label{tab:family}
\end{table*}

\begin{table*}[!H]
\footnotesize
\center
\caption{
{\bf List of reactions of the model of the Bcl--2 family of proteins.}
Reactions no. 46 -- 61 denote production of the correspondent species, 
while the reaction no. 62 denotes degradation of all the species of the model.
}
\begin{tabular}{llll}
\hline
\textbf{No.} & \textbf{Reaction} & \textbf{Forward rate} & \textbf{Reverse rate} \\
\hline
1	& tBid + Bax $\rightarrow$ tBid + aBax	& ka	&	\\
2	& tBid + Bak $\rightarrow$ tBid + aBak	& ka	&	\\
3	& Bim + Bax $\rightarrow$ Bim + aBax	& ka	&	\\
4	& Bim + Bak $\rightarrow$ Bim + aBak	& ka	&	\\
5	& Mcl1 + Puma $\rightleftharpoons$ Mcl1$\sim$Puma	& ks	& km	\\
6	& Mcl1 + Bim $\rightleftharpoons$ Mcl1$\sim$Bim		& ks	& km	\\
7	& Mcl1 + Noxa $\rightleftharpoons$ Mcl1$\sim$Noxa	& ki	& km	\\
8	& A1 + Hrk $\rightleftharpoons$ A1$\sim$Hrk		& ki	& km	\\
9	& A1 + Bik $\rightleftharpoons$ A1$\sim$Bik		& ki	& km	\\
10	& A1 + tBid $\rightleftharpoons$ A1$\sim$tBid		& ks	& km	\\
11	& A1 + Puma $\rightleftharpoons$ A1$\sim$Puma		& ks	& km	\\
12	& A1 + Bim $\rightleftharpoons$ A1$\sim$Bim		& ks	& km	\\
13	& A1 + Noxa $\rightleftharpoons$ A1$\sim$Noxa		& kw	& km	\\
14	& BclXl + Hrk $\rightleftharpoons$ BclXl$\sim$Hrk	& ks	& km	\\
15	& BclXl + Bik $\rightleftharpoons$ BclXl$\sim$Bik	& ki	& km	\\
16	& BclXl + tBid $\rightleftharpoons$ BclXl$\sim$tBid	& ki	& km	\\
17	& BclXl + Puma $\rightleftharpoons$ BclXl$\sim$Puma	& ks	& km	\\
18	& BclXl + Bim $\rightleftharpoons$ BclXl$\sim$Bim	& ks	& km	\\
19	& BclXl + Bad $\rightleftharpoons$ BclXl$\sim$Bad	& ks	& km	\\
20	& BclXl + Bmf $\rightleftharpoons$ BclXl$\sim$Bmf	& ks	& km	\\
21	& Bcl2 + Hrk $\rightleftharpoons$ Bcl2$\sim$Hrk		& kw	& km	\\
22	& Bcl2 + Bik $\rightleftharpoons$ Bcl2$\sim$Bik		& kw	& km	\\
23	& Bcl2 + Puma $\rightleftharpoons$ Bcl2$\sim$Puma	& ks	& km	\\
24	& Bcl2 + Bim $\rightleftharpoons$ Bcl2$\sim$Bim		& ks	& km	\\
25	& Bcl2 + Bad $\rightleftharpoons$ Bcl2$\sim$Bad		& ki	& km	\\
26	& Bcl2 + Bmf $\rightleftharpoons$ Bcl2$\sim$Bmf		& ks	& km	\\
27	& Bclw + Hrk $\rightleftharpoons$ Bclw$\sim$Hrk		& ki	& km	\\
28	& Bclw + Bik $\rightleftharpoons$ Bclw$\sim$Bik		& ki	& km	\\
29	& Bclw + tBid $\rightleftharpoons$ Bclw$\sim$tBid	& ki	& km	\\
30	& Bclw + Puma $\rightleftharpoons$ Bclw$\sim$Puma	& ks	& km	\\
31	& Bclw + Bim $\rightleftharpoons$ Bclw$\sim$Bim		& ks	& km	\\
32	& Bclw + Bad $\rightleftharpoons$ Bclw$\sim$Bad		& ki	& km	\\
33	& Bclw + Bmf $\rightleftharpoons$ Bclw$\sim$Bmf		& ks	& km	\\
34	& Mcl1 + aBax $\rightleftharpoons$ Mcl1$\sim$aBax	& ks	& km	\\
35	& Mcl1 + aBak $\rightleftharpoons$ Mcl1$\sim$aBak	& ks	& km	\\
36	& A1 + aBax $\rightleftharpoons$ A1$\sim$aBax		& ks	& km	\\
37	& A1 + aBak $\rightleftharpoons$ A1$\sim$aBak		& ks	& km	\\
38	& BclXl + aBax $\rightleftharpoons$ BclXl$\sim$aBax	& ks	& km	\\
39	& BclXl + aBak $\rightleftharpoons$ BclXl$\sim$aBak	& ks	& km	\\
40	& Bcl2 + aBax $\rightleftharpoons$ Bcl2$\sim$aBax	& ks	& km	\\
41	& Bclw + aBax $\rightleftharpoons$ Bclw$\sim$aBax	& ks	& km	\\
42	& BclB + aBax $\rightleftharpoons$ BclB$\sim$aBax	& ks	& km	\\
43	& aBax + aBax $\rightleftharpoons$ aBax$\sim$aBax	& kd	& km	\\
44	& aBak + aBak $\rightleftharpoons$ aBak$\sim$aBak	& kd	& km	\\
45	& aBax + aBak $\rightleftharpoons$ aBax$\sim$aBak	& kd	& km	\\
46	& $\rightarrow$ Hrk	& kpHrk	&	\\
47	& $\rightarrow$ Bik	& kpBik	&	\\
48	& $\rightarrow$ tBid	& kptBid	&	\\
49	& $\rightarrow$ Puma	& kpPuma	&	\\
50	& $\rightarrow$ Bim	& kpBim		&	\\
51	& $\rightarrow$ Bad	& kpBad		&	\\
52	& $\rightarrow$ Bmf	& kpBmf		&	\\
53	& $\rightarrow$ Noxa	& kpNox		&	\\
54	& $\rightarrow$ Mcl1	& kpMcl1	&	\\
55	& $\rightarrow$ A1	& kpA1		&	\\
56	& $\rightarrow$ BclXl	& kpBclXl	&	\\
57	& $\rightarrow$ Bcl2	& kpBcl2	&	\\
58	& $\rightarrow$  Bclw	& kpBclw	&	\\
59	& $\rightarrow$  BclB	& kpBclB	&	\\
60	& $\rightarrow$  Bax	& kpBax		&	\\
61	& $\rightarrow$  Bak	& kpBak		&	\\
62	& (All) $\rightarrow$	& kdeg		&	\\
\hline
\end{tabular}
\begin{flushleft}
\end{flushleft}
\label{tab:reactions}
\end{table*}

\begin{table*}[!H]
\footnotesize
\center
\caption{
{\bf Detailed model -- list of parameters.}}
\begin{tabular}{lll}
\hline
\textbf{Parameter} & \textbf{Default value}	& \textbf{Ref.}	\\
\hline
ks	& $1.0\times10^{-4}$	&	\\
ki	& $1.0\times10^{-5}$	& 	\\
kw	& $1.0\times10^{-6}$	&	\\
ka	& $1.0\times10^{-4}$	&	\\
kd	& $1.0\times10^{-4}$	&	\\
km	& $1.0\times10^{-2}$	&	\\
kdeg	& $1.0\times10^{-3}$	&	\\
kpHrk 	& $1.0$			& 1/6 of 6.0, \cite{hua_effects_2005,dlugosz_bcl-2_2006}	\\
kpBik	& $1.0$			& 1/6 of 6.0, \cite{hua_effects_2005,dlugosz_bcl-2_2006}	\\
kptBid	& $0.1$			& 1/2 of 0.2, \cite{hua_effects_2005,dlugosz_bcl-2_2006}	\\
kpPuma	& $1.0$			& 1/6 of 6.0, \cite{hua_effects_2005,dlugosz_bcl-2_2006}	\\
kpBim	& $0.1$			& 1/2 of 0.2, \cite{hua_effects_2005,dlugosz_bcl-2_2006}	\\
kpBad	& $1.0$			& 1/6 of 6.0, \cite{hua_effects_2005,dlugosz_bcl-2_2006}	\\
kpBmf	& $1.0$			& 1/6 of 6.0, \cite{hua_effects_2005,dlugosz_bcl-2_2006}	\\
kpNoxa	& $1.0$			& 1/6 of 6.0, \cite{hua_effects_2005,dlugosz_bcl-2_2006}	\\
kpMcl1	& $10.0$		& 1/6 of 60.0, \cite{kuwana_bid_2002,kuwana_bcl-2-family_2003}	\\
kpA1	& $10.0$		& 1/6 of 60.0, \cite{kuwana_bid_2002,kuwana_bcl-2-family_2003}	\\
kpBclXl	& $10.0$		& 1/6 of 60.0, \cite{kuwana_bid_2002,kuwana_bcl-2-family_2003}	\\
kpBcl2	& $10.0$		& 1/6 of 60.0, \cite{kuwana_bid_2002,kuwana_bcl-2-family_2003}	\\
kpBclw	& $10.0$		& 1/6 of 60.0, \cite{kuwana_bid_2002,kuwana_bcl-2-family_2003}	\\
kpBclB	& $10.0$		& 1/6 of 60.0, \cite{kuwana_bid_2002,kuwana_bcl-2-family_2003}	\\
kpBax	& $60.0$		& 1/2 of 120.0, \cite{kuwana_bid_2002,kuwana_bcl-2-family_2003}	\\
kpBak	& $60.0$		& 1/2 of 120.0, \cite{kuwana_bid_2002,kuwana_bcl-2-family_2003}	\\

\hline
\end{tabular}
\begin{flushleft}
\end{flushleft}
\label{tab:params}
\end{table*}


\end{document}